# The Effect of Shallow vs. Deep Level Doping on the Performance of Thermoelectric Materials

Qichen Song[1], Jiawei Zhou[1], Laureen Meroueh[1], David Broido[2], Zhifeng Ren[3] and Gang Chen[1,*]

[1]Department of Mechanical Engineering, Massachusetts Institute of Technology, Cambridge, Massachusetts 02139, USA

[2]Department of Physics, Boston College, Chestnut Hill, Massachusetts 02467, USA

[3]Department of Physics and TcSUH, University of Houston, Houston, Texas 77204, USA

It is well known that the efficiency of a good thermoelectric material should be optimized with respect to doping concentration. However, much less attention has been paid to the optimization of the dopant's energy level. Thermoelectric materials doped with shallow levels may experience a dramatic reduction in their figures of merit at high temperatures due to the excitation of minority carriers that reduces the Seebeck coefficient and increases bipolar heat conduction. Doping with deep level impurities can delay the excitation of minority carriers as it requires a higher temperature to ionize all dopants. We find through modeling that, depending on the material type and temperature range of operation, different impurity levels (shallow or deep) will be desired to optimize the efficiency of a thermoelectric material. For different materials, we further clarify where the most preferable position of the impurity level within the band gap falls. Our research provides insights in choosing the most appropriate dopants for a thermoelectric material in order to maximize the device efficiency.

[*] Corresponding author: gchen2@mit.edu

Solid-state thermoelectric devices can be used in power generation by converting heat into electricity[1]. The maximum efficiency of thermoelectric power generator is determined by the dimensionless figure of merit $zT = \sigma S^2 T / (\kappa_e + \kappa_h + \kappa_{bp} + \kappa_l)$, where $\sigma$ is the electrical conductivity, $S$ the Seebeck coefficient, $T$ the temperature, $\kappa_e$ and $\kappa_h$ the unipolar electronic thermal conductivity of electrons and holes respectively, $\kappa_{bp}$ the bipolar thermal conductivity and $\kappa_l$ the lattice thermal conductivity[2]. To optimize the efficiency of a thermoelectric generator, a high $zT$ is preferable. Many efforts have been made to enhance $zT$, including reducing the lattice thermal conductivity[3–5] and optimizing the power factor $\sigma S^2$ [6–8]. Specifically, doping is an indispensable step for the material to have a significant electrical conduction, and selecting dopants has become the essential part in optimizing the thermoelectric performance[9].

Introducing dopants into thermoelectric materials usually creates impurity levels within the material's energy band gap. Impurity levels close to the band edge (either conduction band minimum or valence band maximum) are called "shallow" levels, while those far away from the band edge are called "deep" levels[10]. The fraction of dopants ionized to create electrons or holes depends on temperature and the position of the impurity level measured from the Fermi level. In a thermoelectric material doped with shallow levels, it is expected that at a moderate temperature, the donors (acceptors) are nearly all ionized. However, at a higher temperature, minority carriers emerge due to direct thermal excitation across the bandgap, which increases the bipolar thermal conductivity[11] and reduces the Seebeck coefficient due to cancellation of electron and hole currents[12], thus causing a dramatic reduction of $zT$. A thermoelectric material doped with deep levels, on the other hand, requires high thermal energy and thereby a high temperature to ionize the impurities. Therefore, the minority carriers can only be excited at an even higher temperature, *i.e.*, the deep levels suppress the thermal excitation of minority carriers. For a thermoelectric material under an increasingly large temperature gradient, deep levels delay the undesired bipolar effect to higher temperatures and may help to improve the maximum efficiency. There has been continuous interest in optimizing thermoelectric performance through carefully designing the doping, such as blocking minority carriers by heterostructure barriers[12], preferential scattering of the minority carriers[13], stabilizing the optimal carrier concentration[14]. Experiments show that doping deep levels (indium) in $PbTe_{1-x}S_x$ alloy is an effective way to enhance the thermoelectric efficiency[15]. There has not been, however, a systematic study on the effect on thermoelectric performance of different impurity energy levels.

Deep levels can mitigate the bipolar effect and delay the reduction of $zT$ to higher temperatures. The tradeoff, however, is that deep levels usually provide fewer free carriers than shallow levels at the same dopant concentration. As a result, a higher dopant concentration is required for the deep levels to have a comparable performance with shallow levels. Moreover, an inappropriate dopant concentration of deep levels makes the Fermi level deviate from its optimal position. In this letter, using theoretical modeling, we answer the following question: under what conditions are deep/shallow levels preferred? We find that a material with a narrow band gap working under a large temperature difference, for which the bipolar effect can be induced easily, requires deep levels to maximize the efficiency. Besides, a large material parameter $B$, usually a signature of good thermoelectric performance, also implies a significant bipolar contribution to the total thermal conduction and therefore deep levels are desired.

To have a general understanding of the role of deep levels in thermoelectrics, we take parabolic and isotropic conduction and valence bands, where the band curvature is characterized by single-valley density-of-state effective mass $m_d^*$. Assuming that the effective masses and band degeneracies of electrons and holes are equal, the density of majority and minority carriers is written as[16]

$$\begin{aligned} n &= \frac{N_v}{2\pi^2}\left(\frac{2m_d^* k_B T}{\hbar^2}\right)^{3/2} F_{1/2}(\alpha), \\ p &= \frac{N_v}{2\pi^2}\left(\frac{2m_d^* k_B T}{\hbar^2}\right)^{3/2} F_{1/2}(\beta), \end{aligned} \tag{1}$$

where $N_v$ is the band degeneracy, $\hbar$ the reduced Plank constant, $\alpha = (E_f - E_c)/k_B T$, $\beta = (E_v - E_f)/k_B T$ the reduced Fermi level measured from conduction band edge and valence band edge, respectively. The complete Fermi-Dirac integral of $j$ th order is defined as $F_j(\eta) = \int_0^\infty \frac{\xi^j}{1+e^{\xi-\eta}} \mathrm{d}\xi$, where $\xi$ is the reduced carrier energy $\xi = (E - E_c)/k_B T$, $E$ the energy of carriers and $k_B$ the Boltzmann constant. In this paper, the n-type thermoelectric material is studied. The normalized Fermi level $\alpha$ is obtained by solving the charge neutrality equation $n = p + N_d^+$, where $N_d^+ = N_d \Big/ \left(1 + \beta_d e^{(E_f - E_d)/k_B T}\right)$ is the ionized impurity density, $N_d$ the doping concentration and $\beta_d$ the donor degeneracy ($\beta_d$ is set to be 2). Note that the statistics of the deep levels can be complicated[17], but here we still adopt the effective-mass approximation used in shallow levels to treat deep levels, which has been shown to give reasonable results compared with experiments[18] and serves as a starting point to study the effect of impurity energy levels.

The relaxation time approximation is adopted to solve the linearized Boltzmann transport equation to calculate the transport properties[19]. The major scattering mechanism for carriers is assumed to be acoustic deformation potential scattering[20]. The electrical conductivity can then be expressed as,

$$\sigma = \frac{2}{3} N_v \frac{\hbar C_l e^2}{\pi m_d^* \Xi^2} \left[ F_0(\alpha) + F_0(\beta) \right], \tag{2}$$

where $C_l$ is the average longitudinal elastic modulus and $\Xi$ is the deformation potential constant. The Seebeck coefficient is written as,

$$S = \frac{k_B}{e} \left[ \frac{\alpha F_0(\alpha) - \beta F_0(\beta) - 2\left(F_1(\alpha) - F_1(\beta)\right)}{F_0(\alpha) + F_0(\beta)} \right]. \tag{3}$$

The electronic contribution to thermal conductivity of electrons and holes is equal to,

$$\begin{aligned} \kappa_e &= \kappa^* \left[ 3F_2(\alpha) - \frac{4\left(F_1(\alpha)\right)^2}{F_0(\alpha)} \right], \\ \kappa_h &= \kappa^* \left[ 3F_2(\beta) - \frac{4\left(F_1(\beta)\right)^2}{F_0(\beta)} \right]. \end{aligned} \tag{4}$$

And the bipolar thermal conductivity can be described by,

$$\kappa_{bp} = \kappa^* \frac{F_0(\alpha) F_0(\beta)}{F_0(\alpha) + F_0(\beta)} \left[ \xi_g + \frac{2F_1(\alpha)}{F_0(\alpha)} + \frac{2F_1(\beta)}{F_0(\beta)} \right]^2, \tag{5}$$

where $\xi_g = E_g / k_B T$ is the reduced band gap and $E_g = E_c - E_v$ is the band gap. Here, the physical quantity $\kappa^*$, which has the same units as thermal conductivity [W/mK], is defined to be $\kappa^* = \frac{2}{3} N_v \frac{\hbar C_l k_B^2 T}{\pi m_d^* \Xi^2}$. The value of $\kappa^*$ of selected thermoelectric materials is listed in Table (1). As we see in Eqs. (4)-(5), the parameter $\kappa^*$ scales the magnitude of the electronic thermal conductivity and the bipolar thermal conductivity as well. In our model, the lattice thermal conductivity $\kappa_l$ is assumed to be temperature-independent†. Accordingly, the figure of merit $zT$ is given by,

$$zT = \frac{\left[\alpha F_0(\alpha) - \beta F_0(\beta) - 2F_1(\alpha) + 2F_1(\beta)\right]^2}{1/B(T) + \left[F_0(\alpha) + F_0(\beta)\right]\left[3F_2(\alpha) + 3F_2(\beta) - \frac{4F_1^2(\alpha)}{F_0(\alpha)} - \frac{4F_1^2(\beta)}{F_0(\beta)}\right] + F_0(\alpha)F_0(\beta)\left[\xi_g + \frac{2F_1(\alpha)}{F_0(\alpha)} + \frac{2F_1(\beta)}{F_0(\beta)}\right]^2}. \quad (6)$$

In Eq. (6), the material-dependent dimensionless parameter is written as $B(T) = \kappa^*/\kappa_l$ [21]. This parameter has been widely used to identify materials with good thermoelectric performance, though it does not take into account the bipolar effect due to the lack of a band gap parameter[22]. The transport properties of thermoelectric materials vary greatly with temperature. To accurately compare the effect of shallow levels and deep levels, we calculate the device efficiency under different operating temperature ranges. The maximum efficiency derived from the energy conservation equation is obtained as $\eta_{max} = \eta_c \left(\sqrt{1 + ZT_{eng}\alpha_1\eta_c^{-1}} - 1\right) / \left(\alpha_0\sqrt{1 + ZT_{eng}\alpha_1\eta_c^{-1}} + \alpha_2\right)$ and the definition of $ZT_{eng}$ and $\alpha_i$ can be found in literature[23].

We first study the general trend of the dependence of a thermoelectric material's performance on temperature by doping successively deeper levels. In a narrow-bandgap thermoelectric material, the optimal Fermi level (corresponds to the optimal doping concentration that maximizes *zT* at certain temperature *T*) initially lies below the conduction band edge and then rises above the conduction band edge with increasing temperature in Fig. 1(a). This behavior occurs because at low temperatures a high Seebeck coefficient is preferable, which requires the Fermi level to lie slightly lower than the conduction band, while at higher temperatures, the optimal Fermi level should avoid approaching the middle of the band gap to prevent large concentrations of minority carriers that would reduce $zT$ through the bipolar effect. Starting with the same carrier concentration at 300 K, we want to explore how temperature mediates the dopant ionization for different impurity levels and how the Fermi level evolves with temperature relative to the optimal Fermi level. For the shallow levels (green dash-dot line), at room temperature, most of the donors are ionized. With increasing temperature, the donors all become ionized and thermally excited electron-hole pairs become dominant. Accordingly, in Fig. 1(b), $zT$ drops dramatically after expericiencing a peak value and the efficiency starts to decrease at $T_h$ = 1000 K in Fig. 1(c). In Fig. 1(b), the narrow-bandgap material doped with deep levels (blue dashed line) follows a similar trend compared with the shallow levels, yet the peak $zT$ is shifted towards higher temperature and the value of peak $zT$ is slightly smaller. At room temperature, the deep levels are not fully ionized, so we expect a much lower electrical conductivity as a result of fewer carriers being available. However, at high temperatures, the donors are becoming fully ionized and the electron-hole pairs just start to emerge in the system, which explains the mitigated $zT$ reduction. As a result of the supressed onset of the bipolar effect, the efficiency starts to surpass that for the shallow levels when $T_h$ reaches above 900 K in Fig. 1(c). A similar scenario showing delayed bipolar effect has also been reported in experiment[15]. For the case of doping with deeper levels (black dotted line), the temperature dependence of $zT$ is further weakened such that we do not observe a dramatic drop of $zT$ at high temperature in Fig. 1(b). However, the $zT$ falls below the other two curves. This is because for much deeper levels, a higher doping concentration is required to reach the same carrier concentration at room temperature and the Fermi level rises up into the conduction band as the temperature becomes high since the gradually ionized dopant centers release more and more electrons into the system, shown in Fig. 1(a). When the Fermi level is deep inside the conduction band, the Seebeck coefficient is much smaller, leading to a reduced $zT$ over the whole temperature range. In Fig. 1(c), for the case of doping with deeper levels, the efficiency is the lowest in most temperature ranges and keeps increasing as $T_h$ increases. Whereas for a wide-bandgap thermoelectric material, the shallow levels lead to the highest efficiency over the whole temperature range and introducing deeper levels will only reduce the efficiency, shown in Fig. 1(f), due to the fact that a large band gap intrinsically suppresses the excitation of minority carriers, and the optimal choice of dopant is not deep levels.

---

† In the current formalism, a temperature-dependent $\kappa_l$ indeed changes the figure of merit *zT*, yet makes no notable differences in the relative improvement in efficiency after optimization.

Having discussed the general feature of the effects of dopants with different energy levels on the thermoelectric performance, we now discuss how one should select the dopants from the perspective of their energy levels. Practically, there are two variables that can be controlled: the doping concentration $N_d$ and the dopant element, the latter of which determines the energy level $E_d$ of the dopant. To better characterize these dopant properties, we introduce two corresponding dimensionless parameters, $N_d^* = \ln\left(\frac{2N_d\pi^2\hbar^3}{N_v(2m_d^* k_B T_c)^{3/2}}\right)$ and $\xi_d^* = \frac{E_c - E_d}{k_B T_c}$, where $T_c$ is the cold-side temperature. For a given temperature difference, the thermoelectric efficiency is determined by $N_d^*$ and $\xi_d^*$ as shown in Fig. 2(a). There exists an optimal $\xi_d^*$ corresponding to the maximum efficiency for a fixed temperature difference. The origin of this optimum stems from the fact that either too large or too small $\xi_d^*$ will place the Fermi level far away from the position of optimum efficiency while a suitable one should keep the Fermi level relatively close to this optimum across the whole temperature range. Changing from doping with shallow levels ($E_c - E_d$ = 0.01 eV) to doping with deep levels ($E_c - E_d$ = 0.12 eV), there is a 10 % relative improvement in efficiency, when the highest efficiency of each level is compared.

In Fig. 2(b), the optimal $\xi_d^*$ and $N_d^*$ are presented, from which we firstly notice that $\xi_d^*$ and $N_d^*$ follow the same trend, *i.e.*, a higher value of $\xi_d^*$ requires a higher value of $N_d^*$ to reach higher efficiency, meaning that deeper levels need a higher doping concentration. The non-dimensional parameter $\xi_d^*$ is a measure of the difficulty of donor excitation and $N_d^*$ describes the amount of donors provided in the material. A larger $\xi_d^*$ corresponds to a higher binding energy indicating that it is more difficult to ionize the donors; thus more donors are required to reach the optimal number of free carriers determined by the optimal Fermi level. We also note that deeper levels are preferred under a larger temperature difference. This is due to the fact that bipolar thermal conductivity increases rapidly with rising temperature. To counteract the stronger bipolar effect, it is desirable to have donors with deeper levels that will not ionize until higher temperatures. When the temperature difference is small ($\Delta T <$ 100 K), there is no obvious optimal $\xi_d^*$, as the Fermi level can be tuned to close to the optimal position with both deep levels and shallow levels without introducing a significant bipolar effect.

Fig. 2(c) shows the optimal $E_c - E_d$ as a function of band gap and hot-side temperature with respect to different $\kappa^*$. From the figure, we can conclude that impurities with deep levels are preferable to impurities with shallow levels in narrow-bandgap materials. In wide-bandgap materials, the distance required between an impurity level and the conduction band edge is smaller. If the band gap is even larger, the position of optimal impurity level converges because the band gap is large enough to supress minority carrier excitation. For the same $E_g$ and $T_h$, impurities with deeper levels are more preferable as $\kappa^*$ is larger. In the meantime, when $\kappa_l$ becomes smaller (not shown here), a deeper level is also favorable. This is because either a large $\kappa^*$ or a low $\kappa_l$ means that bipolar thermal conduction will dominate. This indicates that a material with large elastic modulus, large band degeneracy, small effective mass, small deformation potential and low lattice thermal conductivity benefits from deep level doping in particular, which is also consistent with the definition of a good thermoelectric material. In other words, a good thermoelectric material demands more attention in choosing the approriate impurity level.

We wish to examine the effect of deep level doping in real material systems. PbTe is chosen as the example material considering it has a narrow band gap and small effective mass as well as a low lattice thermal conductivity. As can be seen in Fig. 2(d), the green line shows the optimum impurity level of PbTe from the abovementioned simple model. The red line shows the optimum impurity level of PbTe as a function of temperature, obtained by a full model that applies the Kane-band model to calculate band structure, acoustic/optical deformation potential scattering, polar scattering[24], and impurity scattering to calculate mobility[25], Callaway model including phonon-phonon scattering[26] in the lattice thermal conductivity calculation and incorporates the temperature dependence of the effective mass and band gap[24], which does not deviate much from the simple model. As the temperature difference increases, the impurity level required becomes deeper. Indium (black dashed line) has been shown to be a deep level dopant substituting for Pb in PbTe[28], while iodine (blue dashed-dot line) is a shallow level dopant substituting for Te[29]. For the case of high $T_h$, we would expect that indium doped PbTe has a higher efficiency than iodine doped due to the surpression of the bipolar effect. In fact, if phonon-impurity scattering[27] is included in the full model, the relative improvement in efficiency from iodine doped to indium doped can be as high as 11% (6% by simple model), due to the extra benefit that the large mass difference between indium and lead gives a reduced $\kappa_l$.

Although the solubility of impurities is considered in the model, the idea of doping with the optimal level is worth further study and it will be even more powerful to adopt it in a material capable of hosting a considerable amount of deep donors.

In summary, we have studied the effect of depth of an impurity level on the thermoelectric figure of merit $zT$ and efficiency using a two-parabolic-band model and relaxation time approximation. We find that deep levels can improve the thermoelectric performance by delaying the bipolar effect, depending on the material's characteristics including the band gap and material parameter *B*, as well as the operating temperature. The performance of shallow level doped thermoelectric materials at high temperatures is degraded due to the bipolar effect, which can be amended by doping with deep levels. Introducing impurities with deep levels requires a larger doping concentration and those with much deeper levels can diminish $zT$ by pushing the Fermi level into the band. Thus there is an optimal impurity level to maximize the device efficiency. Moreover, when the temperature difference is large and the band gap is small, deep levels are desired to increase the efficiency. For different materials, the optimized position of impurity level varies depending on their transport properties. For a thermoelectric material with a large material parameter *B*, doping with deep level impurities results in much larger benefits. Our results can provide guidance on choosing the most appropriate dopant to enhance the performance of thermoelectric materials.

## Acknowledgments:

This material is based upon work supported by the Solid State Solar-Thermal Energy Conversion Center (S3TEC), an Energy Frontier Research Center funded by the U.S. Department of Energy, Office of Science, Office of Basic Energy Sciences under Award No.

Table I. Properties of selected thermoelectric materials at 300K.

| | $E_g$ (eV) | $\kappa^*$ (W/mK) | $\kappa_l$ (W/mK) | Note |
|---|---|---|---|---|
| PbS[24] | 0.42 | 0.18 | 2.5 | n-type |
| PbSe[24] | 0.29 | 0.28 | 1.6 | n-type |
| PbTe[24] | 0.31 | 0.32 | 2.0 | n-type |
| $Bi_2Te_3$[30] | 0.13 | 0.60 | 1.7 | n-type |
| $Bi_2Se_3$[31–33] | 0.24 | 0.12 | 2.8 | n-type |
| $Sb_2Te_3$[34–36] | 0.24 | 3.18 | 1.0 | p-type |
| SiGe[37] | 0.96 | 1.57 | 8.8 | n-type |
| $Mg_2Si$[38–40] | 0.77 | 9.25 | 7.9 | n-type |
| $Mg_2Ge$[38–40] | 0.74 | 4.40 | 6.6 | n-type |
| $Mg_2Sn$[38–40] | 0.35 | 0.50 | 5.9 | n-type |
| $CoSb_3$[41–43] | 0.22 | 5.43 | 10 | p-type |
| ZrNiSn[44] | 0.51 | 0.89 | 6.0 | n-type |

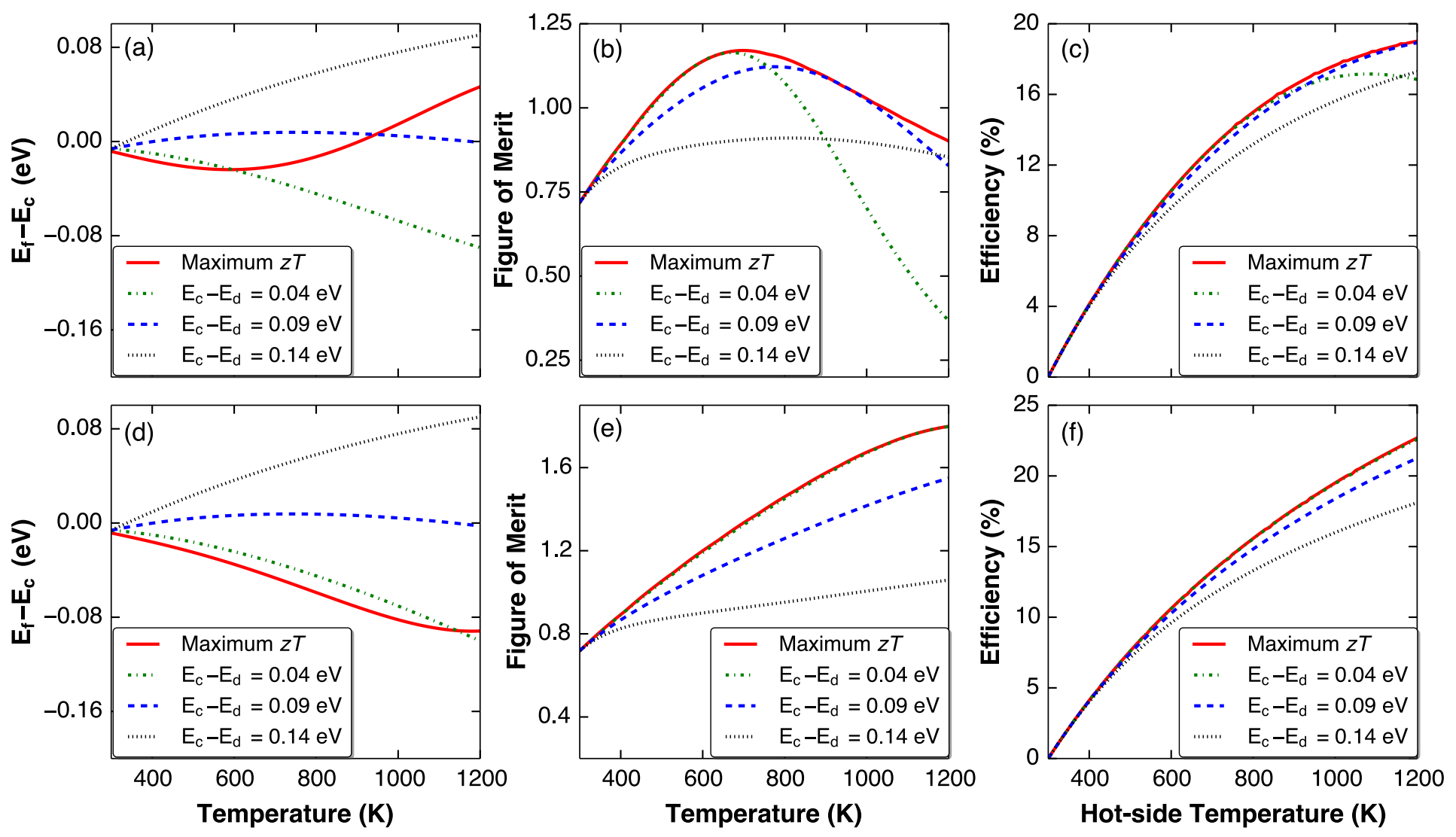

Figure 1: (Color online) The Fermi level and the figure of merit $zT$ of a thermoelectric material as a function of temperature with different impurity level $E_c - E_d$ and the corresponding device efficiency at different hot-side temperature and the same cold-side temperature (300 K). Here, $E_c - E_d$ is defined to be the distance between the impurity level and the conduction band minimum in unit of eV. (a)-(c) show the case of a narrow-bandgap thermoelectric material (0.4 eV) and (d)-(e) show the case of a wide-bandgap thermoelectric material (0.9 eV). In the calculation, $\kappa^* = 0.5\text{W/mK}$, $\kappa_l = 2\text{W/mK}$.

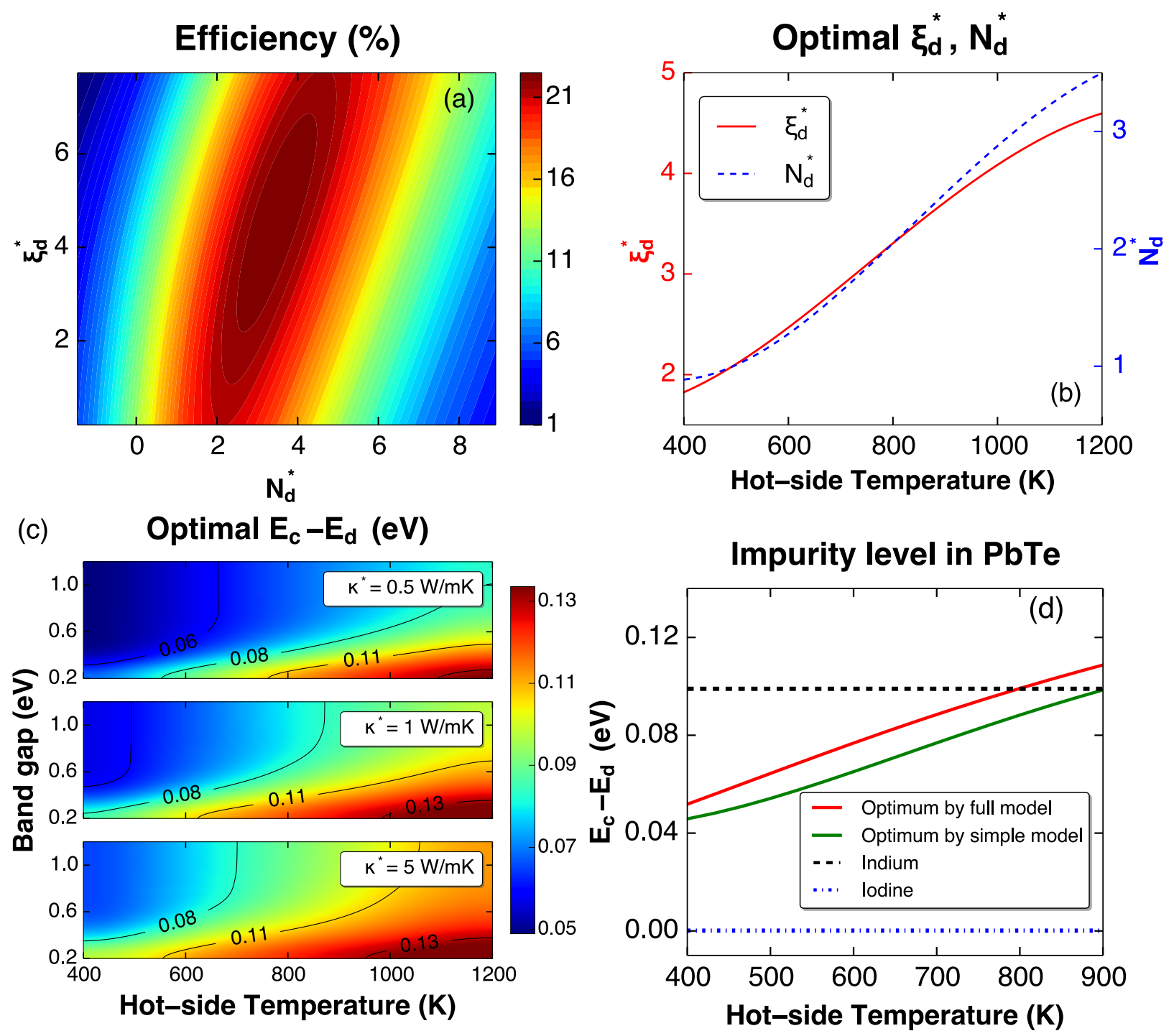

Figure 2: (Color online) (a) Given the cold-side (300 K) and hot-side temperature (1200 K), the efficiency of a thermoelectric material as a function of the dimensionless numbers $\xi_d^*$ and $N_d^*$ (the band gap is 0.4 eV). (b) $\xi_d^*$ and $N_d^*$ required to reach the maximum efficiency at different hot-side temperatures. In the calculation, $\kappa^*$ = 0.5 W/mK, $\kappa_l$ = 1 W/mK. (c) Optimal impurity level $E_c - E_d$ as a function of band gap and hot-side temperature $T_h$ with different $\kappa^*$. In the calculation, $\kappa_l$ = 1 W/mK. (d) Optimal impurity level $E_c - E_d$ of PbTe as a function of hot-side temperature $T_h$, obtained by the simple model (green solid line) and the full model (red solid line). Black dashed line is the impurity level of indium doped PbTe and blue dashed-dot line is the impurity level of iodine doped PbTe. $T_c$ = 300 K and $T_h$ ranges from 400 K to 900K.